\documentstyle{AGILEpsr}

\input epsf.sty
\input epsfig.sty
\input psfig.sty

\def \AAP #1 #2 {{\em Astron. Astrophys.\/} {\bf #1}, #2}
\def \AAL #1 #2 {{\em Astron. Astrophys. Lett.\/} {\bf #1}, L#2}
\def \AAR #1 #2 {{\em Astron. Astrophys. Rev.\/} {\bf #1}, #2}
\def \AAS #1 #2 {{\em Astron. Astrophys. Suppl. Ser.\/} {\bf #1}, #2}
\def \AJ #1 #2 {{\em Astron. J.\/} {\bf #1}, #2}
\def \ANNREV #1 #2 {{\em Ann. Rev. Astron. Astrophys.\/} {\bf #1}, #2}
\def \APJ #1 #2 {{\em Astrophys. J.\/} {\bf #1}, #2}
\def \APJL #1 #2 {{\em Astrophys. J. Lett.\/} {\bf #1}, L#2}
\def \APJS #1 #2 {{\em Astrophys. J. Suppl.\/} {\bf #1}, #2}
\def \APSS #1 #2 {{\em Astrophys. Space Sci.\/} {\bf #1}, #2}
\def \ASR #1 #2 {{\em Adv. Space Res.\/} {\bf #1}, #2}
\def \BAIC #1 #2 {{\em Bull. Astron. Inst. Czechosl.\/} {\bf #1}, #2}
\def \JSQRT #1 #2 {{\em J. Quant. Spectrosc. Radiat. Transfer\/} {\bf #1}, #2}
\def \MN #1 #2 {{\em Mon. Not. R. Astr. Soc.\/} {\bf #1}, #2}
\def \MEM #1 #2 {{\em Mem. R. Astr. Soc.\/} {\bf #1}, #2}
\def \PLR #1 #2 {{\em Phys. Lett. Rev.\/} {\bf #1}, #2}
\def \PASJ #1 #2 {{\em Publ. Astron. Soc. Japan\/} {\bf #1}, #2}
\def \PASP #1 #2 {{\em Publ. Astr. Soc. Pacific\/} {\bf #1}, #2}
\def \NAT #1 #2 {{\em Nature\/} {\bf #1}, #2}
\def \SAIT #1 #2 {{\em Mem.\ Soc.\ Astron.\ It.\/} {\bf #1}, #2}
\def \MESS #1 #2 {{\em The Messenger\/} {\bf #1}, #2}
\def \ASTRNACH #1 #2 {{\em Astron. Nach.\/} {\bf #1}, #2}
\def \AGPSR #1 #2 {{\em ASI Special Publication\/} {\bf #1}, #2}

\begin{opening}

\title{Implications of Low Energy X-ray Emission from Millisecond Radio Pulsars }
\author{M. Ruderman$^{1,2,3}$}
\institute{$^1$Department of Physics, Columbia University\\
$^2$Columbia Astrophysics Laboratory, Columbia University\\
$^3$Institute of Astronomy, Cambridge}
\date{} 
\end{opening}

\begin{document}

\def\lesssim{\mathrel{\hbox{\rlap{\hbox{\lower4pt\hbox{$\sim$}}}\hbox{$<$}}}}
\def\gtrsim{\mathrel{\hbox{\rlap{\hbox{\lower4pt\hbox{$\sim$}}}\hbox{$>$}}}}

\oddpagefooter{}{}{} 
\evenpagefooter{}{}{} 
\medskip  

\begin{abstract} 

Low energy X-ray emission (0.1-10 keV) from all six millisecond radio
pulsars (MSPs) for which such emission has been reported support a proposed
pulsar magnetic field evolution previously compared only to radiopulse data:
old, very strongly spun-up neutron stars become mainly orthogonal rotators
(magnetic dipole moment perpendicular to stellar spin) or aligned rotators.
The neutron star properties which lead to such evolution are reviewed.  Special
consideration is given to agreement between predictions and observed X-ray 
emission for the aligned MSP candidate PSR J0437-4715.

\end{abstract}

\medskip

\section{Introduction}

\def\lesssim{\mathrel{\hbox{\rlap{\hbox{\lower4pt\hbox{$\sim$}}}\hbox{$<$}}}}
\def\gtrsim{\mathrel{\hbox{\rlap{\hbox{\lower4pt\hbox{$\sim$}}}\hbox{$>$}}}}

Properties of the radioemission from the most rapidly spinning disk population
millisecond radio pulsars (MSPs) support the neutron star model-based
expectation that these pulsars should be mainly orthogonal rotators or aligned
rotators. There are three kinds of observational evidence for this.  

1) A large fraction of the MSPs have two radio-subpulses of
comparable strength separated in time by about half a period ( P/2) , a
signature property of an orthogonal rotator (Chen \& Ruderman  1993,
Jayawardhana \& Grindlay 1996). In a sample of 8 of the fastest spinning
disk-MSPs at least 4, and possibly 6, are in this category ( Chen et al 1998);
among canonical radiopulsars only a few percent are ( Lyne \& Manchester 1988). 

2) The well studied original MSP, PSR 1937, is observed to have very
different linear polarization properties in two narrow radio-subpulses an
interval P/2 apart, just as expected from the spin-up evolution of that pulsar
into an orthogonal rotator whose dipole moment is on the neutron star's (NS's)
spin-axis at the NS crust-core interface (CR 93). 

3) At least one pulsar in the chosen MSP octet, PSR J0437-4715, and
possibly several, have only a single very broad, structured radiopulse.  Its
pulse-width is so large ($\sim$270$^{\circ}$ =3P/4) that a plausible description
of the 
observed radiation beam seems to require an almost aligned beam (and dipole
moment) observed from a line-of-sight direction  near  that of the NS's
spin-axis (Gil \& Krawczyk 97).  Because an almost spin-aligned radiobeam from a
spinning NS sweeps out a smaller solid angle on the sky then that same beam
would from an orthogonal rotator, one observation of an aligned MSP out of the
8 in the sample indicates a very considerably larger aligned fraction.  

In Section 2 we consider the significance of MSP low energy X-ray light-curves
when these data are combined with those from the same MSPs' radiopulses.
Section 3 summarizes relevant NS model predictions fro stellar magnetic field
evolution as these stars were spun-up to become MSPs.  Section 4 presents model
predictions for an aligned MSP's thermal X-ray emission areas and compares them
to those inferred from PSR J0437 observations. In no case is there a conflict
between what is expected from  special properties of NSs and what is observed. 

\section{ X-ray and radio light curves}

Low energy (0.1-10 keV) x-ray emission has been reported from 6 MSPs previously
identified by their pulsed radioemission. The 6 X-ray and radio light curves
are shown in Figure 1 (Becker and Aschenbach 2002). The upper curves show those
for X-ray emission, the lower ones for radioemission. There is no significant
difference between the radio and x-ray profiles except  possibly in PSR 0218
(Fig.1 upper middle) and PSR 1821 (Fig.1 upper left). In those two MSPs the
X-ray light curves seem to resolve an ambiguity: their radioemission light
curve profiles contain a sub-peak structure from the double crossing of a
hollow cone beam rather than emission from three truly separate beams.  
Among the 6 MSPs of Figure 1 ,five reinforce support for the orthogonal rotator
interpretation suggested by the radio light curves. One, PSR J0437, clearly
does not. However, its single very broad  X-ray light curve does not contradict
the nearly aligned rotator interpretation for it. Why observed X-ray emission
from a hot polar cap can be so strongly phase-modulated even though the
observer's line-of -sight is near the spin axis is discussed in CRZ 98. 

\begin{figure}
\epsfysize=12cm 
\centerline{\epsfbox{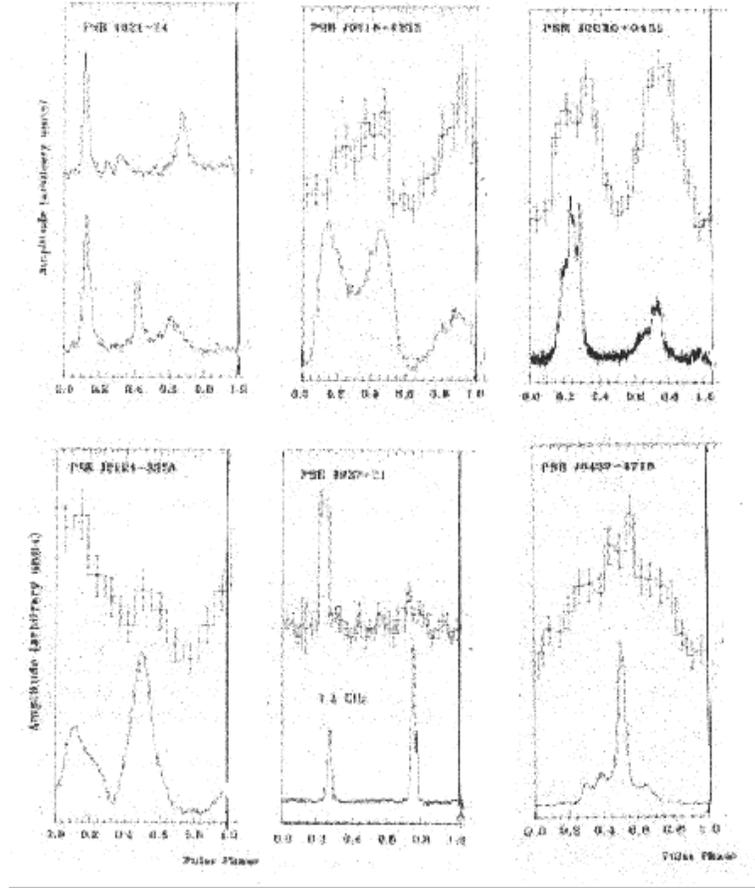}}
\caption[h]{Low energy X-ray light curves for the six radio MSPs for which such
X-ray emission has been reported (Becker \& Aschenbach 2002 ).  Radio profiles
are shown below the X-ray ones. The phase alignments between the X-ray and radio 
profiles have been arbitrarily fitted for PSRs J0030 and J2124. }
\end{figure}

\section{ Neutron star spin-up driven magnetic flux squeezing}

We consider next why MSP genesis based upon very prolonged spin-up of older NSs
with initially nearly canonical dipole moments can lead  to much weaker dipole
moments and ,preferentially, to orthogonal or aligned rotators. 
Soon after its birth a canonical NS is well described (except, possibly, for a
small central region) as an almost uniformly rotating fluid core (radius R = 10 
km) of neutrons (n )together with protons( p) and electrons several percent as
abundant. This core is surrounded by a thin solid crust of thickness $\Delta
\sim$ 1 km. Early in its life (age $< 10^3$ years) the core will have cooled
sufficiently that its neutrons become superfluid (SF-n) and its protons
superconducting (SC-p).  A core of SF-n cannot rotate uniformly. Instead, the
nearly uniform vorticity of 4 $\pi$/P in a rotating normal fluid fragments into a
very dense almost parallel array of quantized vortex lines. The SF-n vortex
number density (per unit area) 

\begin{equation}
%
n_V ~=~ \frac{2 \Omega m_n}{\pi \hbar} ~\sim ~\frac{2~10^4}{P(sec)} ~cm^{-2}
\end{equation}

These vortices are represented by the eight  straight vertical lines in the
Figure 2 cartoon representation of a NS interior. As the NS spins-up (down)
they remain parallel as they move in (out) toward (away from) the NS spin axis
to maintain the vortex line density of Equation 1. The required addition  
(depletion) in total number of vortex lines needed for this comes from the
creation (disappearance) of zero length vortex lines at the spin-hemisphere
equator of the crust-core interface. Within the SC-p superconductor (estimated to be
"Type II")  of the NS core magnetic field fragments into an extremely dense
array of quantized flux-tubes, each containing a magnetic flux  $\Phi _0 = \pi
\hbar c / e \sim 2~10^{-7}$G cm$^2$. One such flux-tube, among the 10$^{30}$ in
the interior of a typical NS, 
is shown in the Figure 2 cartoon. Unlike their neutron vortex line
counterparts flux tubes turn and twist, a relic of a young highly conducting
NS's initially complicated toroidal and poloidal magnetic fields after its
violent birth . The area density of the quantized flux-tubes 

\begin{figure}
\epsfysize=6cm 
\centerline{\epsfbox{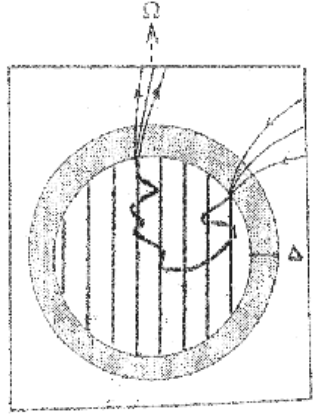}}
\caption[h]{ Interior of a magnetized neutron star with a crust thickness $\Delta$
(not to scale).  The vertical bars symbolize quantized superfluid neutron
vortex lines.  Only a single core magnetic flux-tube is indicated.}
\end{figure}

\begin{equation}
%
n _{\Phi} ~=~ \frac{B}{\Phi _0} ~\sim ~5~10^{18} ~B_{12} ~cm^{-2}
\end{equation}

Within the non-superconducting  crust the flux tubes merge into a
microscopically smooth B. 

MSPs are generally believed to achieve their very high spin from a surrounding
accretion disk fed by a companion (still present around PSR J0437). As the NS
spin-up begins and continues (typically for longer than 10$^8$ years) the
original SF-n vortex array moves inward toward the NS spin-axis. The inward
moving vortices must either cut through the core's magnetic flux array in which
they are embedded or else carry those flux-tubes inward with them. Detailed
calculations show that because of the SF-n-vortices' very slow inward velocity
during this long spin-up ($<$ 10$^{-9}$ cm s$^{-1}$) the inward moving SF-n
vortex array 
enforces a similar inward co-motion of the much denser, more flexible and
complicated, flux-tube array through which it passes (Ruderman et al
1998). Because of motion-induced flux-tube bunching, such forced flux-tube
co-motion would be expected even if the inward velocity of the SF-n vortices
were very much faster (R 01). [In the much earlier spin-down era in the history
of this NS , SF-n vortices moved outward from the NS spin-axis, carrying the
core's flux tubes outward with them. There is strong supporting evidence for
just such magnetic field evolution in much younger spinning-down radiopulsars
(R 01)].   

Figure 3 shows the expected NS surface field evolution during spin-up for three
prototype initial surface field configurations. (The very long time scale for
spin-up to a MSP greatly exceeds the several x10$^6$ year Eddy diffusion time
through the thin stellar crust. Therefore even if movement of an overstressed
crust is neglected, crust conductivity is sufficient to insure that on long
time scales the crust's surface field reflects that of the core surface a
km. below it.)    
In Figure 3 's case ( b)  flux leaving the NS's upper spin-hemisphere returns
to the star in that same hemisphere. In this case prolonged spin-up of the NS
must result in a dipole moment orthogonal to the NS spin-axis , positioned on
that axis where it meets the core's surface (cf. also Figure 4). NS spin-up
from an initial period $P_0$ to a final $P_1$ reduces the NS dipole moment by a
factor $~(P_1 / P_0)^{1/2} ~\sim 10^{-2}$. It is this diminishing dipole strength
that allows 
accretion to achieve such very strong spin-up. 
In Figure 3's case ( a) all returning flux from the upper spin-hemisphere comes
back to the stellar surface in its lower spin-hemisphere. From this initial
configuration prolonged spin-up results in an aligned pulsar. Here, however,
there can only be an increase in the dipole moment component parallel to the
spin. Therefore such a NS could not be spun-up to a very small period MSP
unless its initial dipole moment was already less than 10$^9$ G.  
The Figure 3 case ( c) NS is nearly that of case ( b). Here, however, a small
fraction of the flux out of the upper spin-hemisphere returns to the stellar
surface in its lower hemisphere. During stellar spin-up the large orthogonal
dipole component become much smaller as in case ( b) , but the small aligned
component is not similarly quenched. Whether, after $P_0 \to P_1 ~\ll~ P_0$,  the
strongly 
quenched orthogonal moment or the initially much smaller but unquenched aligned
one dominates depends on both the initial field distribution and on
$P_1 / P_0$. However, when $P1 ~\ll~ P_0$ is achieved,  high fractions of both
orthogonal and aligned MSPs  are expected.  

\begin{figure}
\epsfysize=8cm 
\centerline{\epsfbox{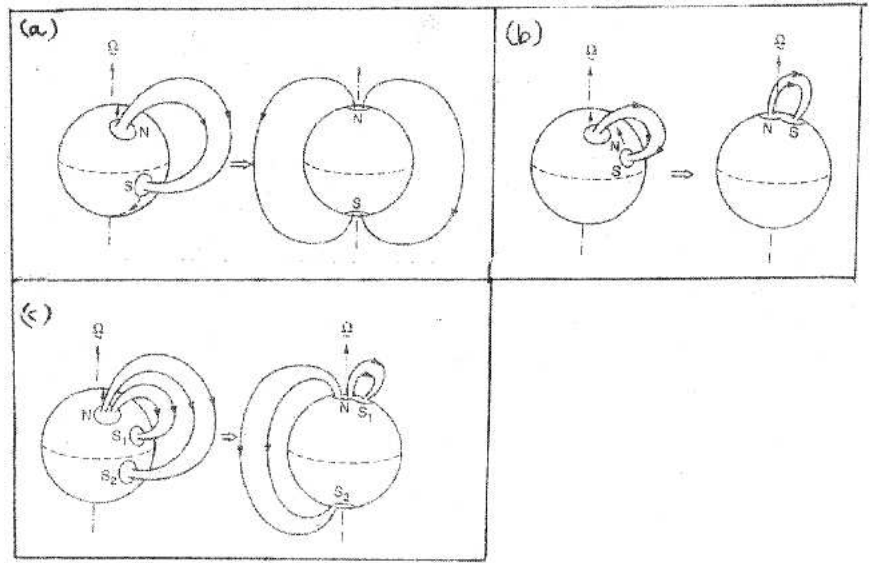}}
\caption[h]{Evolution of the surface magnetic field during prolonged spin-up of
a neutron star. Case ( a ): flux-lines from the upper spin-hemisphere of the
neutron star return to the stellar surface only in the lower spin
hemisphere. Case ( b ): flux-lines from the upper spin-hemisphere return only
to that same hemisphere. Case ( c ): an intermediate case in which some of the
flux out of the upper spin-hemisphere  returns to each. } 
\end{figure}

\section{Aligned MSP polar caps and PSR  J0437}

Figure 4 shows details of final "spin-up squeezed" surface magnetic fields
after NS spin-up to very short period MSPs. Its left panel is that for the
orthogonal dipole after the pre-spin-up field configuration 
of Figure 3 ( b). Agreement between the expected radioemission with this left
panel distribution and radio observations of the MSP orthogonal rotator PSR
1937 has been considered in CR (93). The "spin-up squeezed " structure of the 
surface magnetic field of an aligned pulsar is shown in the middle panel of
Figure 4, and in more detail in the right panel of it. In such a MSP the
surface polar cap is only a short distance (the crust thickness $\Delta$, about 1 km 
above a magnetic pole. One consequence is that the polar cap magnetic field  is
about $(R/ \Delta)^2$ times larger than it in conventional NS models with the
same NS dipole moment  at the center of the star or from a uniformly magnetized
core (CRZ 98). However, the magnetic flux through the polar cap (= the magnetic
flux through the pulsar's "light cylinder") remains the same. Therefore, the
polar cap radius of a strongly spun-up aligned  MSP  (PSR J0437 has P=5.75~ms)
should be  

\begin{equation}
r_1 ~\sim ~\Delta \left( \frac{\Omega R}{c} \right)^{1/2} ~=~ 190 \left
( \frac{P}{6 ms} \right)^{1/2} ~m
\end{equation}

rather than the canonical

\begin{equation}
r_{PC} ~= ~R \left( \frac{\Omega R}{c} \right)^{1/2} ~=~ 1.9 \left
( \frac{P}{6 ms} \right)^{1/2} ~km
\end{equation}

In a very old MSP which long ago stopped accreting the main source for
continuing thermal X-ray emission is expected to be from heating of its polar
caps by backflow onto them of extreme relativistic e$^-$ or e$^+$. These come from
separated  e$^{\pm}$ pairs made within or above an accelerator on the open field
line bundle between the polar cap and a rapidly spinning  aligned MSP's light 
cylinder. The e$^-$ or e$^+$ will be funneled down along the open B-field lines onto
the tiny  radius $r_1$ polar cap. [ At the polar cap the energy of each such
inflowing energetic lepton to the polar cap of PSR J0437 (initially acquired in the 
accelerator) ~ 3 ergs, an energy not sensitive to assumptions about accelerator
parameters (Wang et al 1998). The associated power inflow onto a heated polar
cap is more difficult to estimate because of uncertainty about the e$^-$(e$^+$)
flow rate.]  

\begin{figure}
\epsfysize=6cm 
\centerline{\epsfbox{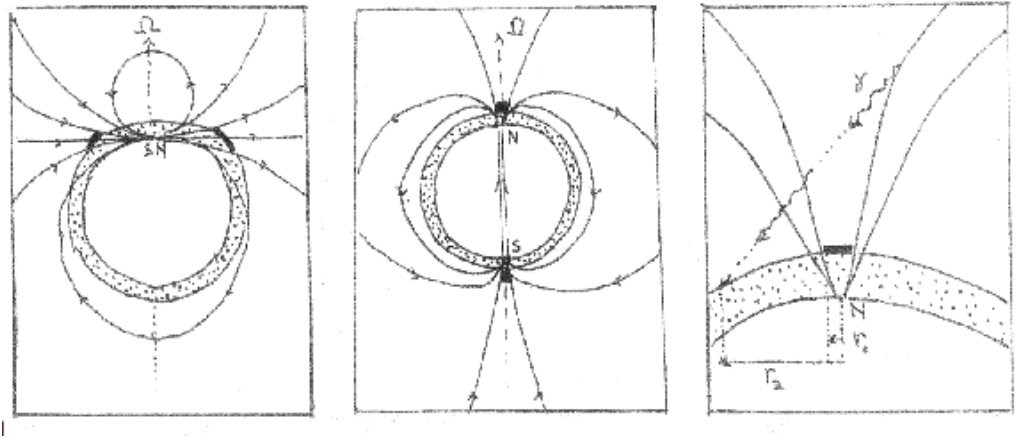}}
\caption[h]{ Magnetic field configurations of strongly spun-up millisecond
pulsars (MSPs). The left panel shows that of a spin-up squeezed orthogonal
rotator. The middle panel shows the spin-up squeezed  field of an aligned
rotator. The right panel is a magnified view of part of the middle one near the
polar cap. Indicated in it is a gamma-ray  of the curvature radiation from an
extreme relativistic inflowing electron (positron) moving along a curved open
field line. The field line curvature and the associated curvature radiation
disappear as the surface is approached.}
\end{figure}

A second source of PSR J0437 surface heating is the 10$^2$ MeV curvature
radiation from its 3 erg leptons as they move down  curved field lines toward
the polar cap. As indicated in the right panel of Figure 4 , when a strongly
spun-up MSP is an aligned rotator these B-field lines are so dominated by the
nearby local pole as they approach the stellar surface that the guiding field
lines become almost straight within a distance ~R of the polar cap. The
inflowing leptons' curvature radiation then drops greatly when they enter that
region. The curvature radiation heating of the aligned MSP's surface, which can
be comparable in magnitude to that from particle flow onto its squeezed
polar cap, is therefore mainly onto a large annulus of inner radius 

\begin{equation}
r_2 \sim r_{PC} \sim 10 r_1 \sim 2 km 
\end{equation}

Thus an aligned PSR J3047 with the spin-up squeezed geometry of Figure 4 should
emit its hottest thermal radiation from a tiny hot polar cap with the radius
$r_1$ of Equation 3 and cooler thermal radiation from a much larger surrounding 
annulus with an inner radius the $r_2$ of Equation 5. 

Observed soft X-ray emission from PSR J0437 fits a combination of a power law
spectrum  together with two black body components of almost equal luminosity
( Zavlin et al 2002, Pavlov et al 2002, B\&A 02). Soft X-ray power law radiation
is expected from the synchrotron emission by e$\pm$ pairs created by
accelerator produced $\gamma$-rays. The thermal components have been interpreted
as emission from a hot polar cap with a radius 
%
$$
r(hot) \sim 120 m
$$

with a cooler rim of radius
%
$$
r(cool) = 2.0 km
$$

These two radii are roughly those of Equations 3 and 5 for the expected spin
squeezed geometry ( right panel of Figure 4) of an aligned PSR J0437.  However,
such agreement as support for this special  B-field geometry is weakened
because a broken power law fit to the whole soft X-ray spectrum has not been 
excluded. If the two black body spectra interpretation is confirmed it would be
a crucial supplement to other evidence for the spin - dipole moment connection
in NS evolution (R 01). 

\acknowledgements

It is a pleasure to thank  K. Chen for his collaborations and W. Becker,
J. Halpern, G. Pavlov, M. Rees, and J. Tr\"umper for generous and informative
conversations.

\end{document}